\documentclass[aps,prb,groupedaddress,amsmath,amssymb,showpacs,
eqsecnum,floatfix,twocolumn]{revtex4}

\usepackage{graphicx}
\begin{document}


\title{The phase diagram of vortex matter in layered superconductors with 
tilted columnar pinning centers}


\author{Chandan Dasgupta}
\email{cdgupta@physics.iisc.ernet.in}
\altaffiliation{Also at Condensed Matter Theory Unit, Jawaharlal Nehru Centre
for Advanced Scientific Research, Bangalore 560064, India}
\affiliation{Centre for Condensed Matter Theory, Department of Physics, 
Indian Institute  of Science, Bangalore 560012, India}
\author{Oriol T. Valls}
\email{otvalls@umn.edu}
\altaffiliation{Also at Minnesota Supercomputer Institute, University of Minnesota,
Minneapolis, Minnesota 55455}
\affiliation{School of Physics and Astronomy,
University of Minnesota, Minneapolis, Minnesota 55455}

\date{\today}

\begin{abstract}
We study the vortex matter 
phase diagram of a layered superconductor in the presence of
columnar pinning defects, {\it tilted} with respect to the normal to the layers.
We use  numerical minimization
of the free energy written as a functional of the time averaged vortex density
of the Ramakrishnan-Yussouff form, supplemented
by the appropriate pinning potential. We study the case where the pin
density is smaller than the areal vortex density. At 
lower pin  concentrations, we find, for  temperatures of the order
of the melting temperature of the unpinned lattice, a Bose glass type
phase which at lower temperatures converts, via a first order transition, to a
Bragg glass, while, at higher temperatures, it crosses over to an interstitial
liquid. At somewhat higher concentrations, no transition to a Bragg glass 
is found
even at the lowest temperatures studied.
While qualitatively the behavior we find is similar to that obtained using the
same procedures for columnar  pins normal to the layers, there are
important and observable quantitative differences, which we discuss.
\end{abstract}


\maketitle

\section{Introduction}
\label{intro}

Equilibrium and dynamic properties of vortex matter in highly anisotropic, 
layered, high-temperature superconductors are known~\cite{review} 
to be strongly affected by the presence
of pinning. Effects of random columnar pinning produced by heavy-ion bombardment
have been studied theoretically~\cite{NV93,Radz95}, experimentally~\cite{Banerjee04} 
and numerically~\cite{TG03,NH04,usprl,us2,us3} for the situation in which both
the columnar pins and the magnetic field are perpendicular to the layers. In
this geometry, if the areal concentration of 
columnar pins exceeds that of vortex lines
(i.e. for $B_\phi > B$, where $B_\phi$ is the matching field and $B$ is the magnetic induction),
the vortex system exhibits a continuous Bose glass (BoG) 
to vortex liquid (VL) transition~\cite{NV93} as the temperature is increased. If, on the
other hand, the relative pin concentration $c \equiv B_\phi/B$ is substantially
smaller than unity, then
the vortex system exhibits a first-order transition~\cite{Radz95,Banerjee04} 
between a high-temperature VL and a 
low-temperature BoG phase that has a polycrystalline structure~\cite{Banerjee04,usprl,us2} 
with grain boundaries
separating crystalline domains of different orientations. The VL into which the BoG melts
has the characteristics of an ``interstitial liquid'' in which some of the vortices remain
localized at the columnar pins, producing solid-like regions around them, whereas the 
remaining, interstitial vortices form liquid-like regions. The pinned vortices delocalize
at a ``depinning crossover'' that occurs at a higher temperature. Numerical 
studies~\cite{NH04,usprl,us2} also indicate the occurrence of a topologically ordered
phase, analogous to the Bragg glass (BrG) phase~\cite{natt,GLD95} in systems with random point pinning,
at low temperatures if the relative concentration of columnar pins is sufficiently small. 

The question of how the behavior described above is modified in the case of tilted
columnar pinning, where the magnetic field is tilted away from the direction of the columnar
pins, has also received considerable attention in the past. Theoretical studies~\cite{NV93,Hwa93}
have considered the geometry in which the columnar pins are perpendicular to the layers and
the applied magnetic field makes an angle $\theta$ with the normal to the layers. These 
studies predict that for $B < B_\phi$, the effects of the correlated nature 
of the columnar pins become less
pronounced as the angle $\theta$ is increased. Specifically, the vortex lines are predicted to
remain locked to the columnar pins if $\theta$ is sufficiently small, producing a ``transverse
Meissner effect''. For larger values of the angle $\theta$, the vortices hop from one columnar
pin to the next one, forming a staircase structure. As $\theta$ is increased further, the 
directional effect of columnar pinning is lost and the vortex lines follow the field 
direction. The low-temperature BoG phase phase persists for small values of $\theta$, but
disappears as $\theta$ is increased beyond a critical value. 
Some of these theoretical predictions have been verified in experiments~\cite{Ammor04}.

The behavior of vortex systems with tilted columnar pinning and $B > B_\phi$ ($c<1$) has been
investigated recently in experiments~\cite{zeldov07} and simulations~\cite{zeldov07,gl07}.  
The experiments were performed on a sample of ${\rm Bi_2Sr_2CaCu_2O_8}$
(BSCCO) with a small concentration
of random columnar pins tilted at an angle of 45$^\circ$ from the normal ($z$-direction) to the
copper oxide layers. The magnitude and direction of the applied magnetic field $\bf H$ were 
varied and the location of the BoG to interstitial VL transition in the $H_z$ versus
temperature ($T$) plane was determined for several values of the tilting angle $\theta$ between  
the directions of the magnetic induction $\bf B$ and the columnar pins. The values of $\bf H$
considered in the experiment were such that the number density of pancake vortices on the
layers (determined by $B_z$) is higher than that of the columnar pins. The main result of
this experiment is that the temperature at which the BoG to VL transition occurs for a fixed
value of $H_z$ is  {\it independent} of the tilt angle $\theta$. The temperature at which
the inhomogeneous VL (called ``vortex nanoliquid'' in Ref.\onlinecite{zeldov07}) crosses over
to the depinned, homogeneous liquid was also found to be independent of $\theta$ for a fixed
value of $H_z$. The simulations were performed for a fixed number density of pancake vortices
on the layers (fixed $B_z$) and different orientations of the columnar pins, keeping the
number density of pinning centers on each layer fixed at a value lower than that of 
pancake vortices. Both Josephson and electromagnetic interactions between pancake vortices
on different layers were included in the simulations. The results of the simulations were
found to be consistent with the experimental observation that the locations of the   
thermodynamic transitions are independent of the angle between the columnar pins and
the applied field if the number densities of pancake vortices and pinning centers on each
layer (i.e. the values of $B_z$ and $B_\phi \cos\psi$ where $\psi$ is the angle between
the layer normal and the direction of the columnar pins) are held fixed.

These results are surprising because tilting the columnar pins away from the direction of
the layer normal introduces ``frustration'' in the system in the following sense. If the
pinning potential of each pinning center is sufficiently strong and the temperature sufficiently
low (these conditions are satisfied in the experiment and simulation described above), then
nearly all the pinning centers on each layer would be occupied by pancake
vortices. For columnar pins perpendicular to the layers,
the pinned vortices on different layers would then be
aligned directly on top of one another. This alignment of the pancake vortices 
in the direction of
the layer normal minimizes both the Josephson and electromagnetic interactions between
vortices on different layers. However, if the columnar pins are tilted away from the layer
normal, then the pinned pancake vortices on different layers would not be aligned directly on top
of one another, thereby increasing the energy associated with the interlayer
interactions of these vortices. For $B_z > B_\phi \cos\psi$ (the case considered in 
Refs.~\onlinecite{zeldov07,gl07}), interstitial pancake vortices that are not localized at 
pinning centers can relieve this frustration to some extent by forming a staircase-like
structure in which they remain aligned in the direction of the layer
normal for a few layers and then shift in the direction of the tilt. This, however, would
increase the energy associated with the interaction of pancake vortices on the same layer
because the positions of the interstitial vortices relative to those of the pinned ones,
which shift in the direction of the tilt by a constant amount as one goes from one 
layer to the next one, would not be optimal on all the layers. Thus, tilting the columnar
pins away from the direction of the layer normal should increase the frustration arising
from the competition between the interaction of the vortices with the pinning centers and
the intervortex interactions. This should have a measurable effect on the transition 
temperatures unless the energy associated with interlayer interactions among 
the pancake vortices is  negligibly
small compared to the other energy scales (the intralayer interactions and the pinning 
energy) of the problem. Since increased frustration tends to lower the temperature at which
an ordering transition occurs, the transition temperatures of the vortex system are expected
to decrease as the tilting angle is increased from zero.

To shed some light on this problem, we have studied the structural and thermodynamic
properties of a system of pancake vortices in a strongly anisotropic, layered superconductor
in the presence of tilted columnar pinning, using a mean-field, free-energy based numerical
method developed in our earlier studies~\cite{usprl,us2,us3,prbv,dv06,dv07} of vortex matter with
different kinds of pinning. In this method, the free energy of a system of pancake vortices
interacting among themselves and with pinning centers is written as a functional of the 
time-averaged local areal density of the vortices. Only the electromagnetic interaction 
between pancake vortices on different layers is considered. Different phases, represented by
different local minima of the free energy,  
are obtained by numerically minimizing the free energy, starting from different initial
configurations of the local density. In this description,
a first order phase transition between two phases corresponds to a
crossing of the free energies of two distinct minima representing the two
phases. Here, we use parameters appropriate for  BSCCO
and fix the areal density of pancake vortices at a value corresponding to 
$B_z$ = 2 kG
for the component of the magnetic induction normal to the layers. 
This corresponds to the experimental situation where the applied magnetic field $\bf H$
is in the $z$-direction and its magnitude is such that $B_z$ equals 2 kG.
We
consider different concentrations of columnar pinning centers, keeping their areal density 
smaller than that of the pancake vortices, so that the relative pin concentration 
$c \equiv B_\phi \cos\psi/B_z$
is much smaller than unity. The columnar nature of the pins is modeled
by repeating the positions of the pinning centers on successive layers with a constant
shift in the case  of tilted pins.  We then compare the results obtained for tilted pins 
with different
tilting angles with those obtained for the same in-plane 
configuration of pinning centers, but without
any tilt (without any shift for pins oriented in the direction
of the layer normal) to analyze the effects of tilting the columnar pins. The main results of our study
are summarized below.

The structural and thermodynamic properties of the systems with
tilted columnar pins are found to be very similar to those found in our earlier
studies~\cite{usprl,us2} of vortex systems in which both the magnetic field and a  
small concentration of random columnar pins are perpendicular to the layers. Specifically,
for small values of the relative pin concentration $c$ defined above, we find,  at 
low temperatures, two distinct 
minima of the free energy. At both these minima, nearly all the pinning 
centers are occupied by vortices, and both the pinned and the interstitial vortices form
lines that are tilted in the direction of the columnar pins. The degree of alignment of the 
vortices in the direction of the tilt is nearly perfect. 
One of these two minima corresponds to the BoG phase in which the vortices on each layer
exhibit substantial short-range translational and bond-orientational order, but topological
defects such as dislocations are present in small concentrations. The other minimum is
almost perfectly crystalline over the length scale of our finite samples and exhibits features
characteristic of the topologically ordered BrG phase of systems with weak point pinning. 
At temperatures close to the melting temperature of the vortex system without any pinning,
the BoG phase is the thermodynamically stable one with lower free energy. As the temperature
is decreased, the free energy of the more ordered phase crosses that of the BoG phase at a first
order phase transition, so that the BrG-like phase becomes the thermodynamically
stable one at low temperatures. The minimum representing the BoG phase evolves continuously to the
high-temperature, depinned VL as the temperature is increased -- we do not find a 
first-order transition to the VL for the pin concentrations considered in this study. Using a
criterion based on percolation of liquid-like regions~\cite{us2,us3}, we define a crossover
temperature for the transformation of the BoG to the interstitial VL. This crossover
occurs at a temperature higher than the melting temperature of the vortex lattice in pristine
samples without any pinning. For larger values of
the relative pin concentration $c$, the low-temperature BrG-like phase is absent and 
only the crossover between the BoG and VL phases is found.

Although the general behavior found for tilted columnar pins is qualitatively similar to
that of systems with ``vertical'' columnar pins normal to the layers, 
a detailed comparison between the results
for the same vortex system with tilted and vertical columnar pins with the same 
in-plane arrangement
of the pinning centers
reveals, in contrast with
some previous studies,\cite{zeldov07,gl07} 
a  significant differences between the two cases. First, the temperature of the 
first-order transition between the BrG and BoG phases for small values of $c$
is found to be lower by over 5\% (about one degree) in the case of tilted pins. The temperature of the
BoG to VL crossover for tilted pins is also decreased by a similar amount from 
that for vertical columnar pins. Thus, the expected reduction in the transition temperatures
due to increased frustration in the tilted pin case is observed in our calculation. Second, the
degree of localization of the pancake vortices, measured by the heights of the local density
peaks that represent vortex positions at the free-energy minima, is always slightly lower
when the pins are tilted. This is true for both the vortices trapped at the pinning centers
and the interstitial ones. This is a consequence of the additional tilting-induced 
competition between the pinning potential
and interlayer vortex interactions mentioned above. This competition makes the pinning 
centers less effective in trapping vortices
and reduces the extent of in-plane order by decreasing the degree of localization of the 
interstitial vortices. 

The rest of the paper is organized as follows. The model considered  and the numerical
methods used in our study are described in section~\ref{methods}. The results obtained 
in this study are described in
detail in section~\ref{results}. We conclude in  section~\ref{summary} 
with a discussion of our main results in the context
of those of earlier studies.

\section{Model and Methods}
\label{methods}


The general method we use here is that of minimizing a mean-field free energy
functional with respect to the time-averaged local vortex density $\rho_n(\bf
r)$, where $\bf r$ is a two dimensional vector denoting a location
in the layer $n$. The free energy includes both intrinsic and pinning
terms:
\begin{equation}
F[\rho]=F_{RY}[\rho]+F_p[\rho].
\label{fe}
\end{equation}
For the first term, we take the Ramakrishnan-Yussouff\cite{ry} form: 

\begin{widetext}
\begin{equation}
\beta F_{RY}[\rho] = \sum_{n}\int{d {\bf r}\{\rho_n({\bf r})
\ln (\rho_n({\bf r})/\rho_0)-\delta\rho_n({\bf r})\} } 
 -(1/2)\sum_m \sum_n \int{d {\bf r} \int {d{\bf r}^\prime
C_{mn}({|\bf r}-{\bf r^\prime|}) \delta \rho_m ({\bf r}) \delta
\rho_n({\bf r}^\prime)}} ,
\label{ryfe}
\end{equation}
\end{widetext}
where $\beta$ is
the inverse temperature and the integrals
are two-dimensional. This free energy is defined with respect
to that of a  vortex liquid with uniform density  
$\rho_0 = B_z/\Phi_0$ where $B_z$ is the component of the
magnetic induction in the direction ($z$ direction) normal to the layers 
and $\Phi_0$  the superconducting flux quantum.
In the above expression,
$\delta \rho_n ({\bf r})\equiv \rho_n({\bf r})-\rho_0$ is the
deviation of  ${\rho_n(\bf r})$ from $\rho_0$ and 
$C_{mn}(r)$ is the direct pair correlation
function of the layered vortex liquid~\cite{hansen}
at density $\rho_0$. 
This static correlation function depends on the layer separation $|m-n|$
and on the distance $r$ in the layer plane,
and it
contains all the required information about the interactions in the system.
As in previous work on 
columnar pins\cite{usprl,us2,us3} normal to the layers, we use here the $C_{mn}(r)$ obtained from a 
calculation~\cite{menon1} via the 
hypernetted chain approximation~\cite{hansen} for parameter values appropriate
for  the layered material BSCCO.  
Within these premises, two material parameters enter
the calculations: the London penetration depth $\lambda(T)$ and the
dimensionless parameter $\Gamma$:
\begin{equation}
\Gamma = \beta d \Phi^2_0/8 \pi^2 \lambda^2(T).
\label{gamma}
\end{equation}
where $d$ is the interplanar distance. We will
take here values appropriate
to BSCCO; thus $d=15 \AA$.

The second term in the right side of Eq.~(\ref{fe}) is the pinning term and
we write it in the form:
\begin{equation}
F_p[\rho]= \sum_n \int{d {\bf r} V^p_n({\bf r}) [\rho_n({\bf r})-\rho_0] },
\label{pinpot1}
\end{equation}
where the pinning potential $V^p_n({\bf r})$ is computed by summing over
the positions ${\bf R}_{j,n}$ of the $j$th pinning center in the $n$th plane:
\begin{equation}
V_n^p({\bf r})=\sum_{j} V_0(|{\bf r}-{\bf R}_{j,n}|),
\label{pinpot}
\end{equation}
The potential $V_0$ corresponding to a single pinning center
is taken to be of the usual truncated parabolic form:\cite{daf}
\begin{equation}
\beta V_0(r)=-\alpha\Gamma[1-(r/r_0)^2]\Theta(r_0-r)
\label{single}
\end{equation}
where $r_0$ is the range. In terms of our unit of length $a_0$, defined
by $\pi a_0^2 \rho_0=1$, we take $r_0=0.1 a_0$. For the strength $\alpha$,
which is a dimensionless number, we take the value ($\alpha=0.05$) at
which\cite{prbv} each pinning center pins slightly less than
one vortex in the temperature range studied. This is the same value used
in the previous  studies\cite{usprl,us2,us3} of vertical columnar pins. 
The number of vortices is
determined by $B_z$ and we will consider here a fixed value $B_z$ =2 kG. 
As in the numerical studies of Refs.~\onlinecite{zeldov07,gl07}, the magnitude
and direction of the applied magnetic field $\bf H$ do not appear explicitly
in our calculation. The situation we consider here may be realized experimentally
by applying a magnetic field in the $z$-direction and adjusting its magnitude
to yield the value of 2 kG for the $z$-component of the magnetic induction 
$\bf B$ in the superconductor.  
The pinning columns make an angle $\psi$
with the $z$ direction.  The relative pin
concentration $c$ is (equivalently with the definition given
above) the ratio of the number $N_p$ of columnar pins to the number 
$N_v$ of vortices in the system.

To study the phase diagram we discretize the position variable and
numerically minimize the free energy with respect to the discrete set
of variables $\rho_{n,i}$ where the index $i$ denotes a position in the $n$
layer of the discretized triangular lattice. We have
$\rho_{n,i} \equiv \rho_n({\bf r_i}) A_0$  where
$A_0= h^2 \sqrt 3/2$ is the area of the in-plane computational
cell of lattice constant $h$. 
The computational lattice is of size $N^2 \times N_L$.    
As in previous work\cite{usprl,us2,us3,prbv,dv06,dv07}
we take $h=a/16$ where $a = 1.99 a_0$ is the equilibrium value\cite{prbv} of the lattice
constant of the system in the absence of pinning at the chosen value of
$B_z$. The minimization procedure we use\cite{cdo} ensures the non-negativity
of the variables $\rho_{n,i}$.

There are some computational issues in solving this problem which must
be explained here. We wish to consider the case where the pin concentration
$c$ is much smaller than unity. 
We also want to consider values of the tilt angle $\psi$ in the reasonable
experimental range. The value of $N$ must be large enough so that the number
of vortices present is not too small. The value $N_L$ of the number
of layers in the computational lattice has to  be\cite{dv06} at least several
hundred. There are of course computational limitations: in our recent\cite{dv07}
work on point pinning the total number of 
computational lattice sites attainable was $N_C=N^2N_L=2^{23}$.
But the main problem here is that the periodic boundary conditions in the $z$
direction impose, computationally, an effective ``quantization condition''
on the values of  $\psi$ that can be used and, indirectly,
on the range of $c$ that can be studied. This occurs
for the following reason: implementation
of  periodic boundary conditions is only possible
if, after $N_L$ layers,
the pinning potential repeats itself. Assume that the potential due
to one of the tilted columnar pins is such
that after an integer number $n$ of layers it has shifted horizontally
by another integer $m$ of in-plane computational lattice sites. 
The two  integers $n$ and $m$ determine the tilt angle via 
$\tan \psi =(m h/nd)$. In order to implement the 
periodic boundary conditions in the $z$ direction, the total horizontal
shift (in units of $h$)  after $N_L$ layers, which is $(N_L /n)m$,
has to equal $N$ so that $(N_L/n)=(N/m)$. Thus one also has 
$\tan \psi = (N h/N_Ld)$. This implies, since $h/d\approx 70/15$
for the chosen value of $B_z$, 
that one needs
a large value of $N_L$ in order to keep $\psi$ from being too large. But
one cannot increase $N_L$ arbitrarily, since the total number of computational
sites $N_C$ must remain 
within feasible bounds. The value of $N_L$ must nevertheless be taken
as large as possible, but, given $N_C$ and $N_L$, one must still have a 
number of vortices $N_v=(N/16)^2$ 
large enough. One has to note also that the
value of $N_v$ puts a lower bound on the values of $c$ that can be studied,
since after all one cannot put less than one pin
in the system.  Thus a complicated series
of compromises must be made to optimize the parameter
values for which data are obtained.

With the above in mind, the data presented here have been obtained with
$N_L=1024$. Two values of $N$ have been used: most of
the data have been obtained for $N=96$ (which means
$N_C=2^{20}3^2> 2^{23}$) and additional results
will be presented for $N=128$ ($N_C=2^{24}$). In the first case $N_v=36$
and we have taken $N_p=4$ or $c=1/9$ while  in the second case 
$N_v=64$ and we have taken 
$N_p=8$ and the somewhat larger concentration $c=1/8$. The
number of vortices in our samples is larger that that used in other computational
work.\cite{gl07} At $N=96$ therefore, we have $\tan \psi=0.437$ while at 
$N=128$ we have a larger angle, $\tan \psi=0.583$.  

\section{Results}
\label{results}

We can now discuss the results obtained using the methods described above. 
The accuracy of these procedures has been repeatedly discussed in previous 
work\cite{usprl,us2,us3} and this issue and other technicalities need not be further elaborated
upon here. 
The iteration process continues until the system reaches a local free energy 
minimum. The structure of the system at that minimum is then inferred by 
analyzing the vortex density structure, i.e., the set of variables $\{\rho_{n,i}\}$. One 
needs some initial condition to start the minimization procedure. If one starts
with perfectly disordered initial conditions, ($\delta \rho_{n,i}\equiv 0$) and 
one quenches to a sufficiently high temperature, one obtains a disordered 
minimum structure. The resulting values of
$\{\rho_{n,i}\}$ can then be used as the initial condition set at a nearby
$T$. Ordered structures can be then obtained upon  cooling the system to a 
lower $T$ sufficiently slowly. Ordered states can also be obtained by using a 
crystalline structure (we take that which minimizes 
the pinning energy with 
respect to all the symmetry operations of the lattice) as the
initial configuration. These ordered 
configurations can then be warmed up and of course, they eventually become 
disordered. In general, the ordered configurations are to be identified, as
we will explain below, with 
BrG states while the disordered ones are BoG at lower $T$, becoming eventually 
liquid upon warming. At certain temperatures, more than one local 
minimum may be found, and the values of the free energy then establish which is 
the stable configuration and which are only metastable.

\subsection{Structure of minima at $c=1/9$}

We have studied three random pin configurations at $c=1/9$, $N=96$ (as explained 
above). The behavior for all three configurations is extremely consistent.
For each pin configuration, we have studied also, for comparison purposes,
the behavior of the system with the same pin configuration in the top layer but
with the pinning columns being normal to the layers, that is, parallel to 
the $z$ crystal axis, instead of tilted ($\psi=0$). 
In so doing, we consider the same pin configuration 
at the same value
of $N$ to avoid sample to sample variation or 
finite size effects tainting the comparison.\cite{old} 
The value 
of $N_z$ is immaterial for vertical columns, since the problem should be quasi 
two-dimensional in this case, but we have explicitly verified that the results 
do not change
when $N_z$ is reduced from 1024 to eight. 

It is important and very useful 
to visualize the structure of the free energy minima from
the values of the variables $\{\rho_{n,i}\}$. One 
way of doing so is by considering  the {\it vortex lattice} itself, as opposed to
the computational lattice. From the $\{\rho_{n,i}\}$ set
of values, we can locate the position of a vortex at site $i$ in the $n$ layer 
if the value of $\rho_{n,i}$ at that site is larger that the value of 
$\rho_{n,j}$ at any site $j$ within a distance $a/2$ of site $i$. 
The position of 
these locations can then be directly plotted. This allows a clear visualization of 
the arrangement of the vortices at different minima of the free energy.

\begin{figure} 
\includegraphics [scale=0.5]{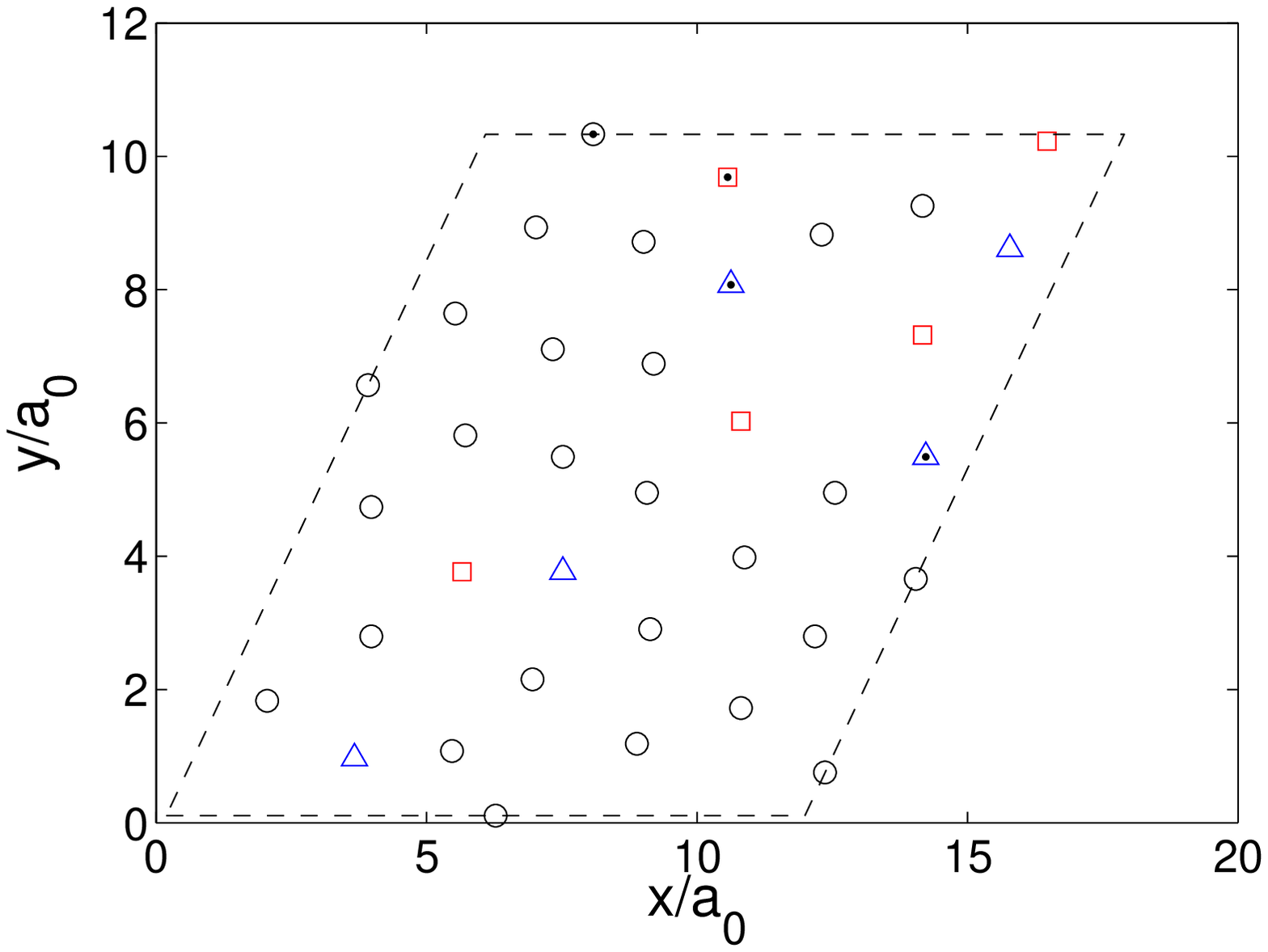}
\includegraphics [scale=0.5] {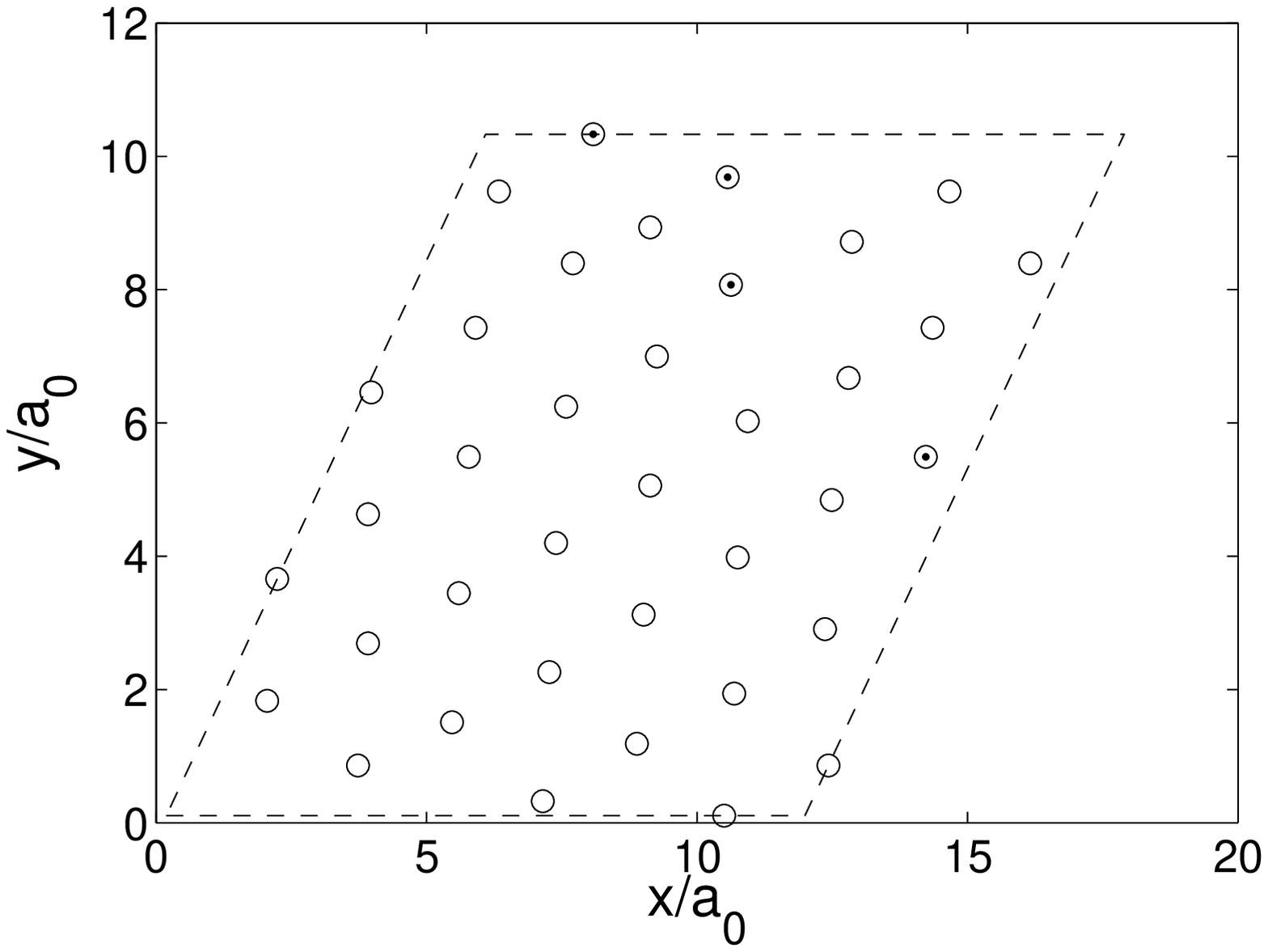}
\caption{(Color online) Vortex lattice structure for the 
BoG (top panel) and BrG (bottom panel) phases. 
The temperature is $17.8 K$ where both
the phases are locally stable. 
Each plot shows vortex positions on 32 different 
layers, appropriately shifted to compensate for a tilt along the pinning 
columns (see text). Dots represent pin
positions, all other symbols are vortex positions. 
The Voronoi analysis (see text) 
of the vortex structure is  shown
by the symbol shape
(and color).  The (black) circles represent ordinary six-fold
coordinated sites, (blue) triangles: 5-fold coordinated, (red) 
squares: 7-fold coordinated. 
}
\label{fig1}
\end{figure}

We first address the question of the degree of alignment of the vortices along the
tilted columnar pins. In our samples, the pin locations shift in the $x$-direction 
by $N=96$ spacings of the computational lattice across $N_L=1024$ layers. Therefore, 
there is a shift of 3 spacings of the computational lattice after every 32 layers.
If the vortices are aligned with the columnar pins, then their positions would
also shift by 3 spacings of the computational lattice after every 32 layers. We 
can check whether this happens by showing in the same plot the vortex positions
on layers $n_l = k+32l$ where $k$ is an arbitrary integer between 1 and 32 and 
$l=0,1,...,31$. To compensate for the expected shift due to the presence of the tilted
columns, we shift the vortex positions on layer $n_l$ by $3l$ spacings of the 
computational lattice in the negative $x$-direction. Then, the plotted vortex positions
after the shifts on all the different layers, $l=0,1,...,31$, for any $k$ should lie on top of
one another if the vortices are aligned with the tilted pins. In Fig.\ref{fig1}, we
show two such plots for two distinct local minima of the free energy at $T=17.8$K.
As discussed below (see Fig.~\ref{fig6}) in detail, 
these two minima correspond to the BoG (top panel)
and BrG (bottom panel) states at a temperature close to the transition temperature
at which their free energies cross. We emphasize that each plot shows the vortex
positions on 32 different layers, corresponding to  $l=0,1,...,31$, shifted appropriately
to compensate for a tilt in the direction of the columnar pins. The dots
in the plot are the pin positions. All other symbols are
vortex lattice sites, the precise meaning of their shapes and colors
is explained below. The vortex positions
on these different layers are found to fall directly on top of one another after the
shifts in both panels of Fig.~\ref{fig1}, so that only
one symbol per site can be seen. This observation indicates that the vortices
are almost perfectly aligned in the tilt direction. 

\begin{figure}
\includegraphics [scale=0.5] {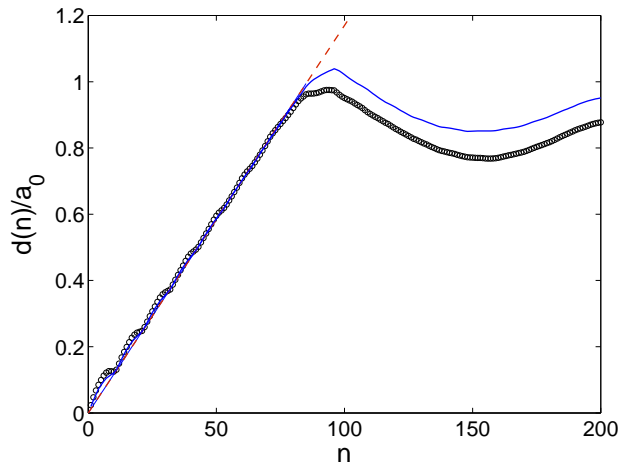}
\caption{(Color online) Distance $d(n)$ between a lattice point and
it nearest neighbor in an  adjacent plane (see text) plotted
vs layer separation $n$ for the BrG and BoG phases at $T=17.8 K$. 
The circles are our results for the BoG minimum, the (blue) solid line 
represents the results for the BrG minimum, and the (red) dashed straight line 
shows the result expected for 
$d(n) \lesssim a_0$ (see text)
for perfect alignment with the tilted columnar pins.}
\label{fig2}
\end{figure}

To examine the degree of alignment of the vortices with the tilted pins for other
values (not multiples of 32) of layer separation, we consider the quantity $d(n)$
which is defined as
the average distance between a vortex site and its nearest neighbor 
in an adjacent plane separated by $n$ layers. This is plotted in Fig.~\ref{fig2} as a 
function the separation $n$ between planes for the same
BrG and BoG minima and temperature as in Fig.~\ref{fig1}. 
If the vortex lines are perfectly tilted, then, from the geometrical 
considerations in Sec.~\ref{methods} and the numerical
values given there, it follows that 
a plot of $d(n)$ vs. $n$ should be a straight line with slope $s=(Nh)/N_L \simeq 0.01165 a_0$
for smaller values of $n$. Departure from a straight line is to be expected if
$n$ exceeds the value for which $d(n)$ reaches a value close to $a_0$, since
$a_0$ is (as previously mentioned) approximately
half of the average spacing $a$ 
between nearest-neighboring vortices on a layer. This is because for such 
larger values of $n$, the vortex
in layer $(n+m)$ that has the smallest lateral separation from a vortex on layer $m$ is {\it not}
the one located at a position shifted by $ns$ in the direction of the tilt from the position 
of the vortex in layer $m$. Thus, since $d(n)$ measures the smallest lateral separation 
between two vortices
located on planes separated by $n$ layer spacings, the linear increase of $d(n)$ with $n$ should
be observed only for $d(n) \lesssim a_0$ or 
$n \lesssim a_0/s \approx 86$.
One can see from Fig.~\ref{fig2} that
a straight line with the expected slope fits the results perfectly well in the relevant
$n$ range and this, together with the argument in the previous paragraph shows that 
as stated in the Introduction, the 
vortex lines are indeed 
nearly perfectly tilted along the direction of the pinning columns. 
This behavior is a consequence of the dominance of the pinning energy over interlayer
vortex interactions for the realistic parameter values used in our calculation.

Turning now to the structure
in the $xy$ plane, we have analyzed the structure of the vortex arrangement in each plane
by means of a Voronoi construction. A Voronoi construction in any lattice 
is performed by dividing it into cells, one cell
per lattice point, each cell consisting of the region 
of space which is closest to a certain lattice point than to any other. For
a crystalline lattice, this is the Wigner-Seitz cell. In general, the 
number of sides of the Voronoi cell surrounding a lattice point 
is the number of neighbors of the lattice
point. The Voronoi analysis then reflects directly the defect structure. 
The use of different symbols in Fig.~\ref{fig1} is 
meant to show examples of such Voronoi plots for the shifted lattice. 
We see that from the point of view of the Voronoi construction there is a
contrast 
between the two cases shown, at the same $T=17.8 K$ where two phases are locally stable
and  have approximately the same free
energy. The state in  the top panel contains a considerable number of defects,
as can be seen by the adjacent site pairs with five or seven neighbors,
while the state in the bottom panel contains none. Hence the first state
can at least tentatively 
be identified as a BoG state while  the phase in the bottom 
panel, which in the spatial scale of the computation looks like
a perfect crystal, can be identified as a BrG with a more
ordered structure than the BoG. 

\begin{figure}
\includegraphics [scale=0.5] {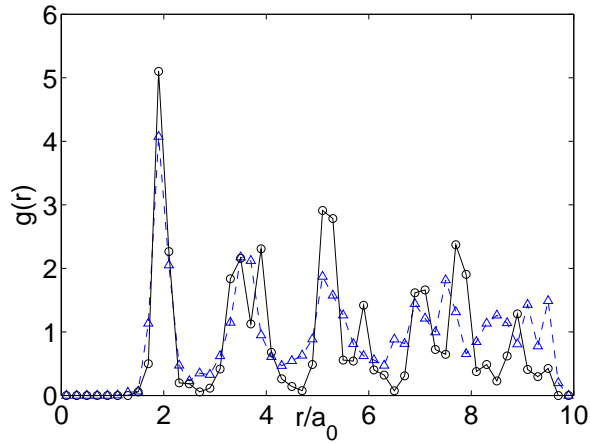}
\caption{(Color online) Angularly averaged in-plane correlation function $g(r)$.
Results for the same states and $T$ as in Fig.~\ref{fig1} are shown.
The (black) line and circles are for the more ordered (BrG) state and the (blue) line
and triangles are for the BoG. }
\label{fig3}
\end{figure}

One can alternatively describe the structure and verify the above
identifications by studying the density 
correlation functions. It is straightforward to extract from the 
vortex positions the in-plane angularly averaged two-point correlation function 
$g(r)$ of the vortex positions,
defined as
\begin{equation}
g(r)= \frac{A}{N_L N_v (N_v-1)} \frac{\sum_n \sum_{i\ne j} m(n,i) m(n,j) f_{ij}(r,\Delta r)}
{2 \pi r \Delta r},
\label{gofr}
\end{equation}
where $m(n,i) = 1$ if the computational lattice site $i$ on layer $n$ corresponds to a vortex
position (i.e. if the local density peaks at this computational lattice site), and $m(n,i)=0$
otherwise, $A$ is the area of the sample in the $xy$-plane, and $f_{ij}(r,\Delta r) =1$ if the distance
between the lattice sites $(n,i)$ and $(n,j)$ lies between $r$ and $r+\Delta r$ (we use $\Delta r=0.2 a_0$ in
our calculation), and $f_{ij}(r,\Delta r)=0$ otherwise. The normalization of 
$g(r)$ is
such that it should approach unity in the large-$r$ limit if there is no long-range
translation order in the planes. For a perfect triangular lattice, the first 5 peaks of $g(r)$
should occur at $r=1.99a_0$, $3.44a_0$, $3.98a_0$, $5.26a_0$ and $5.97a_0$. This $g(r)$ is {\it different} 
from the more familiar pair distribution function that measures the two-point 
correlation of the local density. In particular, information about the degree of
localization of the local density peaks corresponding to the vortex
positions is not contained in $g(r)$ because only the positions
of these peaks are used in its calculation. 
Examples of $g(r)$ are plotted in Fig.~\ref{fig3} for the 
same two cases as in Fig.~\ref{fig1}. Again, we see the contrast
between the two cases. Although the relatively small size of the system precludes
studying the very long $r$ behavior, one can see that the correlation function
for the state which in Fig.~\ref{fig1} exhibited
no defects 
has a more ordered structure 
(higher and better defined peaks at the values of $r$ for which sharp peaks are
expected for a triangular lattice) than the
one we tentatively identified as a BoG state based on the Voronoi constructions
of Fig.~\ref{fig1}. Thus, this analysis if $g(r)$  confirms the identifications made
based on direct visualization and the Voronoi construction.

\begin{figure}
\includegraphics [scale=0.5] {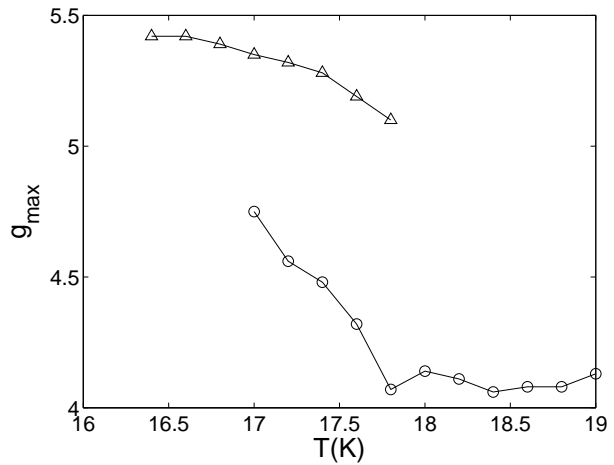}
\caption{ The quantity $g_{max}$ (see text)  used
as a measure of the order parameter,
plotted as a function of $T$ for both the BoG and
the BrG phases. Triangles: ordered (BrG)  state, Circles: BoG.}
\label{fig4}
\end{figure}

Next, in Figure \ref{fig4} we consider a measure of the order as a
function of temperature. There are a number of ways in which one can define an
``order parameter'' and here we choose the value of $g(r)$ at its first $r>0$
peak. This quantity, which we call $g_{max}$, 
is plotted as a function of $T$ for the same configuration
presented in the previous figures. We
do this for both the BrG phase and
the BoG one. As we shall see below in the discussion associated with
Fig.~\ref{fig6}, the BrG does not exist, even as a metastable, state for $T>17.6$
and the same holds for the BoG at $T<17.0$, hence the ranges plotted. We see
that this quantity decreases with $T$ in either case but that it is considerably
larger in the BrG than in the BoG, as one would expect. At $T\approx 17.5 K$ where, as
we shall see below, the free energies of the two states cross, there is a marked
discontinuity in the equilibrium value of $g_{max}$. The nearly constant value of $g_{max}$
for the BoG phase at temperatures higher than 17.8K is a reflection of the above
mentioned fact that the $g(r)$ considered here does not take into account 
the broadening of the local density peaks with increasing temperature.

\begin{figure} 
\includegraphics [scale=0.5]{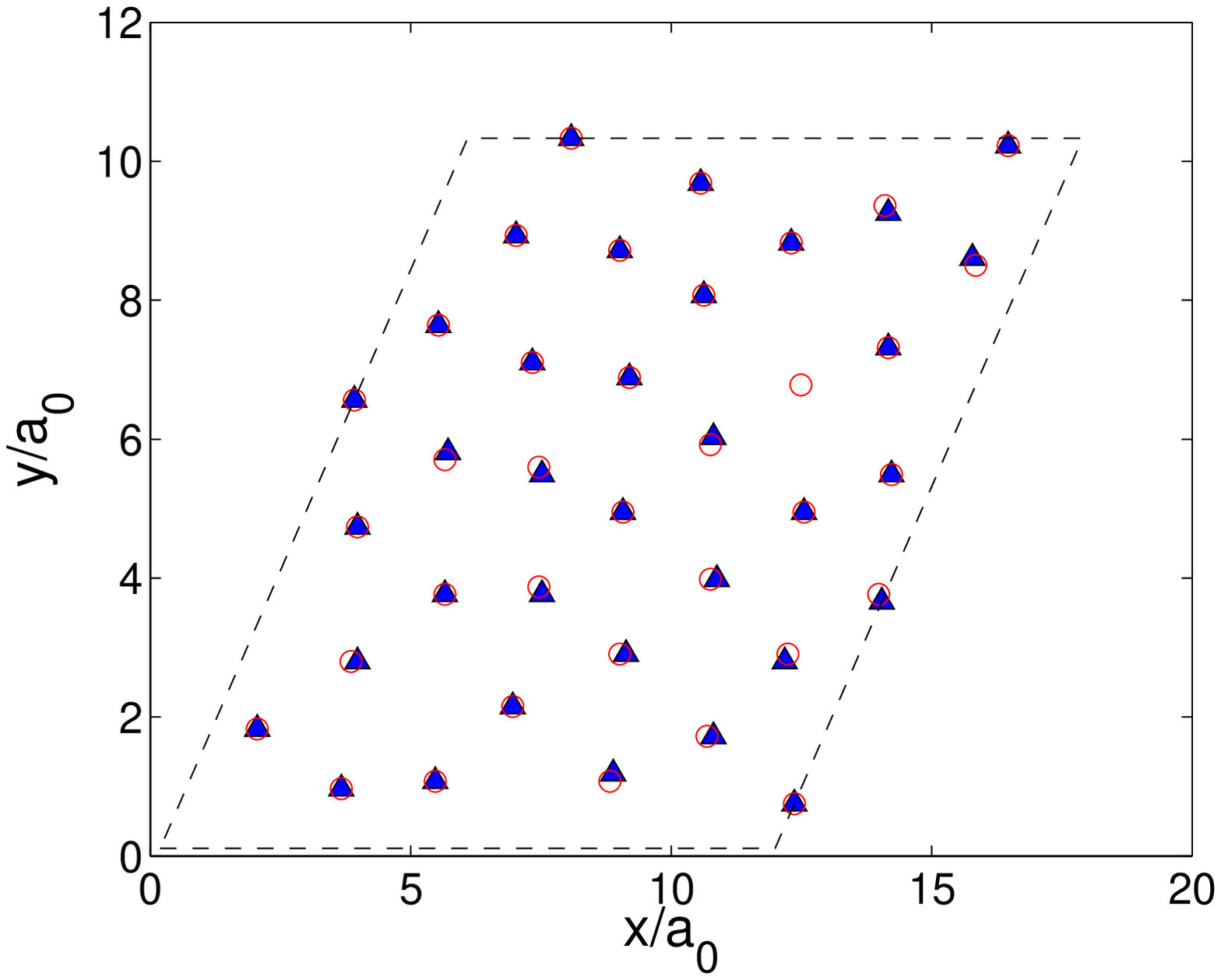}
\includegraphics [scale=0.5] {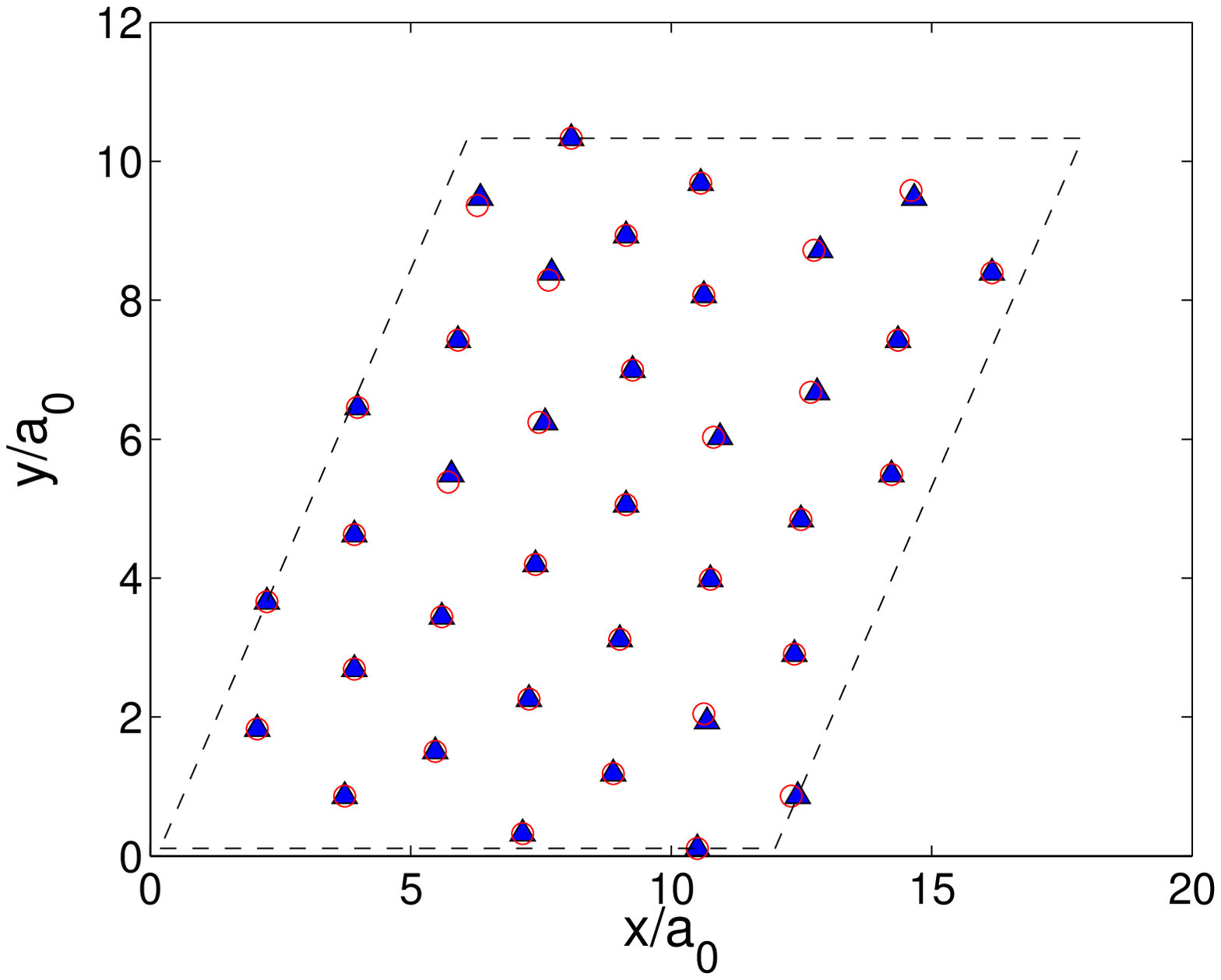}
\caption{(Color online) Comparisons of 
the in-plane structures of BoG and BrG minima obtained (see text)
for tilted and vertical pins at the same
temperature and same in-plane pin configuration. Top panel: BoG phase at $T=18.4$K,
(red) circles: vertical pins, (blue) triangles: tilted pins. Bottom panel: same
for the BrG phase at $T=17.8$K.}
\label{fig5}
\end{figure}

We end this section with a comparison of the structures of the BoG and BrG minima
obtained for the same in-plane pin configuration, but for tilted pins in one case and 
vertical pins
in the other case. In Fig.~\ref{fig5}, we show vortex position plots similar to those in 
Fig.\ref{fig1} except that no Voronoi analysis is performed. 
The top panel shows the data for the BoG phase at $T=18.4$K and the bottom panel shows the
results for the BrG phase at $T=17.8$K. It is clear from these plots that the in-plane
structure for tilted and vertical pins are very similar in both the BrG and BoG phases.
The degree of alignment with the pins is also found to be very similar for tilted and 
vertical pins. There are differences, however, between the vertical and tilted cases, as
we will see below. 

\subsection {Free energy and phase transitions}
\begin{figure}
\includegraphics [scale =0.6]{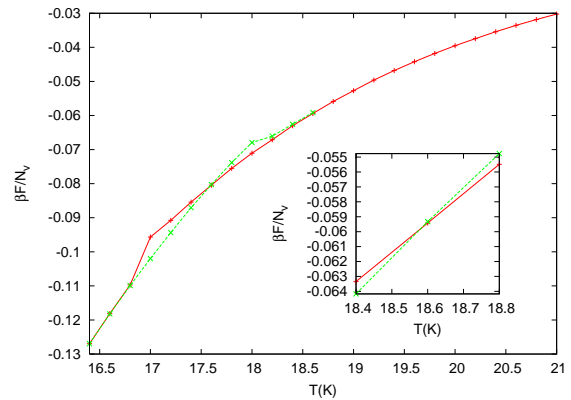}
\caption{(Color online) Free energy vs temperature. 
The data points are the
results, the lines join the data points. The (green) $\times$
symbols are for the BrG state and the (red) plusses for
the BoG. The main plot shows the free energy per
vortex for a tilted configuration at $c=1/9$ (see text). The inset shows the
same data, in a restricted region, for the same configuration but vertical
pinning lines.}
\label{fig6}
\end{figure}

The minimization procedure yields, of course, the value of the free energy at
each local minimum. By considering the free energy values as a function of $T$
the possible phase transitions in the system can be studied. 
In Fig.~\ref{fig6} we show typical results at $c=1/9$. The main plot is for the
tilted case with $N=96$ in which case, as explained above, $\psi=0.41$. The
free energy per vortex is plotted as a function of temperature. At high
temperatures only one state is stable. The corresponding free energy is plotted
as the (red) crosses. By analyzing the results
at each $T$  as explained in the previous 
subsection, we find that this state is disordered, a  BoG. It exists down to
$T=17.0$K, where, as one can see in  the figure, it becomes unstable to the other
state. This other state, the free energy of which is denoted by the (green) 
$\times$ signs connected by dotted lines, is 
found in the same way to be the BrG state. At temperatures 
in the range $17 K\le T \le 18.2 K$ both states can be found, one being of course
only metastable. The crossing of the free energies occurs at $T \approx 17.6 K$  where
therefore a {\it first order} transition occurs, as seen  by the difference in slopes
of the free energy and the discontinuity of the order parameter in Fig.~\ref{fig4}.

The insert shows, in a reduced temperature range, similar results for the same
pin configuration but at $\psi=0$ (vertical pins). We see that in that case the
first order transition occurs near $T=18.6 K$, about one degree higher than in the
tilted case. This one degree shift occurs
for all pin configurations investigated at this value
of $c$: although the values of the individual transition temperatures 
show some sample-to-sample variation, the one degree shift
always occurs. We see then that for $c=1/9$ increasing the angle $\psi$ leads to a
notable decrease of the temperature at which the  BrG transforms to the BoG.

\begin{figure}
\includegraphics [scale=0.5] {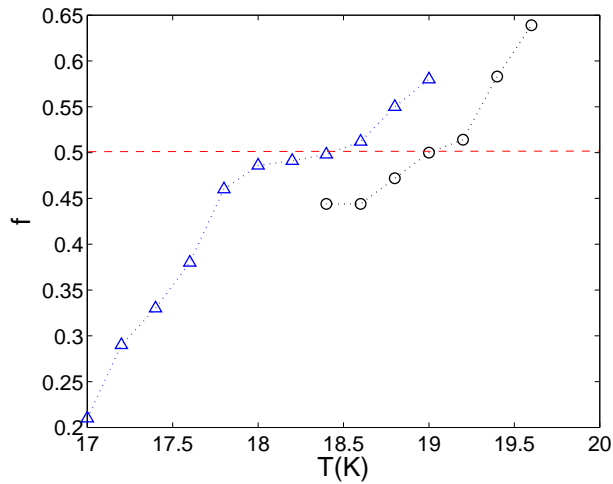}
\caption{Fraction $f$ of liquid-like sites for BoG minima, plotted as a function
of $T$. The triangles are for the tilted pin system, the circles for
the same configuration but with vertical pins.}
\label{fig7}
\end{figure}

At higher temperatures the BoG crosses over to an interstitial liquid phase. As
we have seen in the vertical pin case\cite{usprl,us2,us3} this transition coincides with the 
onset of percolation of the liquid phase. The determination of this transition is
shown in Fig.~\ref{fig7}. The quantity plotted there is the fraction $f$ of the liquid-like
local density peaks as a function of temperature. A vortex lattice site is assumed to
be liquid-like if\cite{us2} the local value of $\rho_{n,i}$ does not exceed
$3\rho_0$ (excluding of course the pinning sites). This fraction of liquid-like sites
is small at lower $T$ and it rises rapidly up to temperatures higher than the first order
transition. Then it flattens somewhat and it crosses the value of $1/2$ (the threshold
value for site percolation on a triangular lattice) at a higher
temperature $T\approx 18.4$K. We take this to be the temperature of crossing over
from the BoG to the IL region. In Fig.~\ref{fig7} results are also plotted for the
vertical pins case. The percolation crossover is found to occur at a slightly higher
temperature, $T \approx 19.0$K, for vertical pins.

\begin{figure}
\includegraphics [scale=0.5] {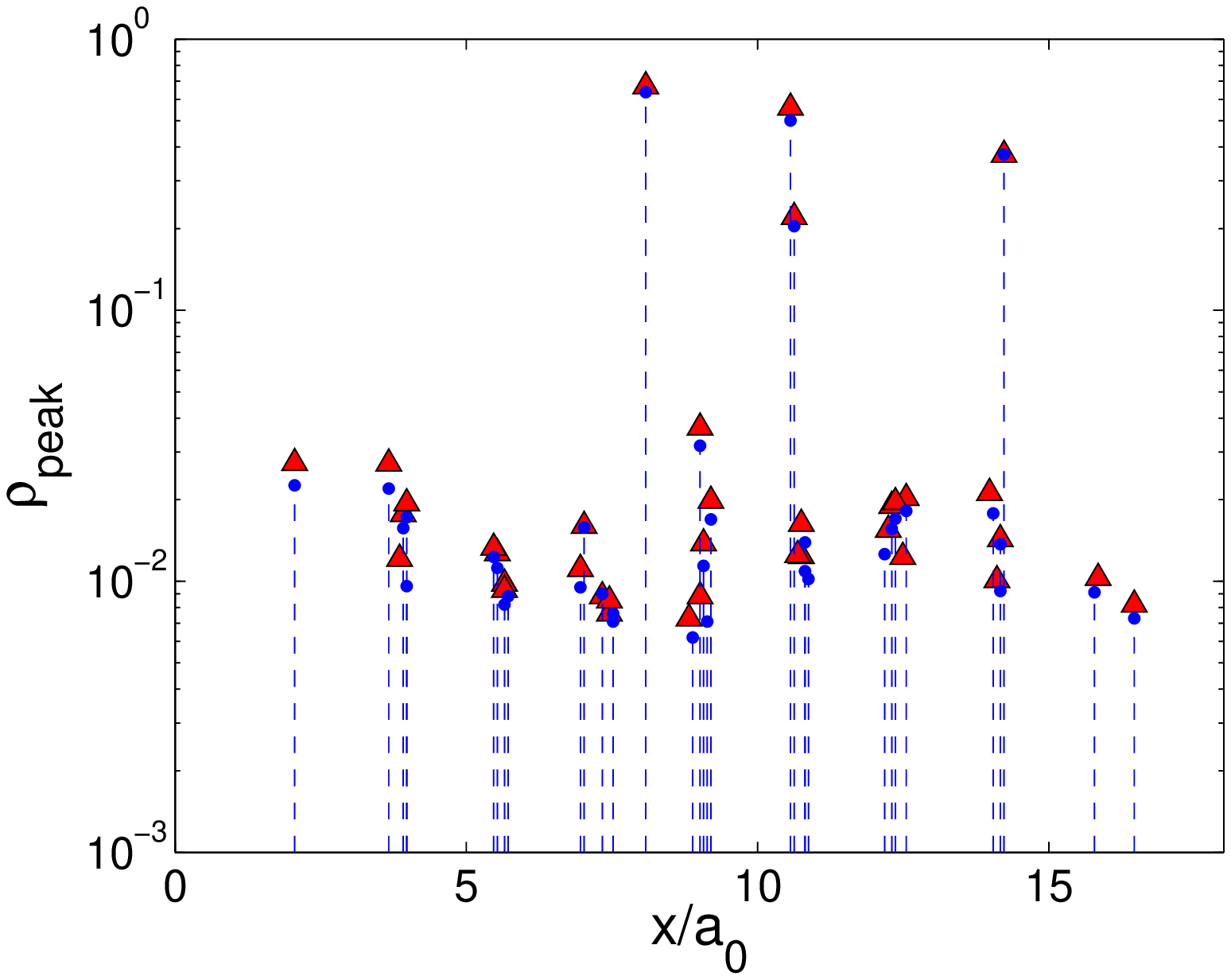}
\includegraphics [scale=0.5] {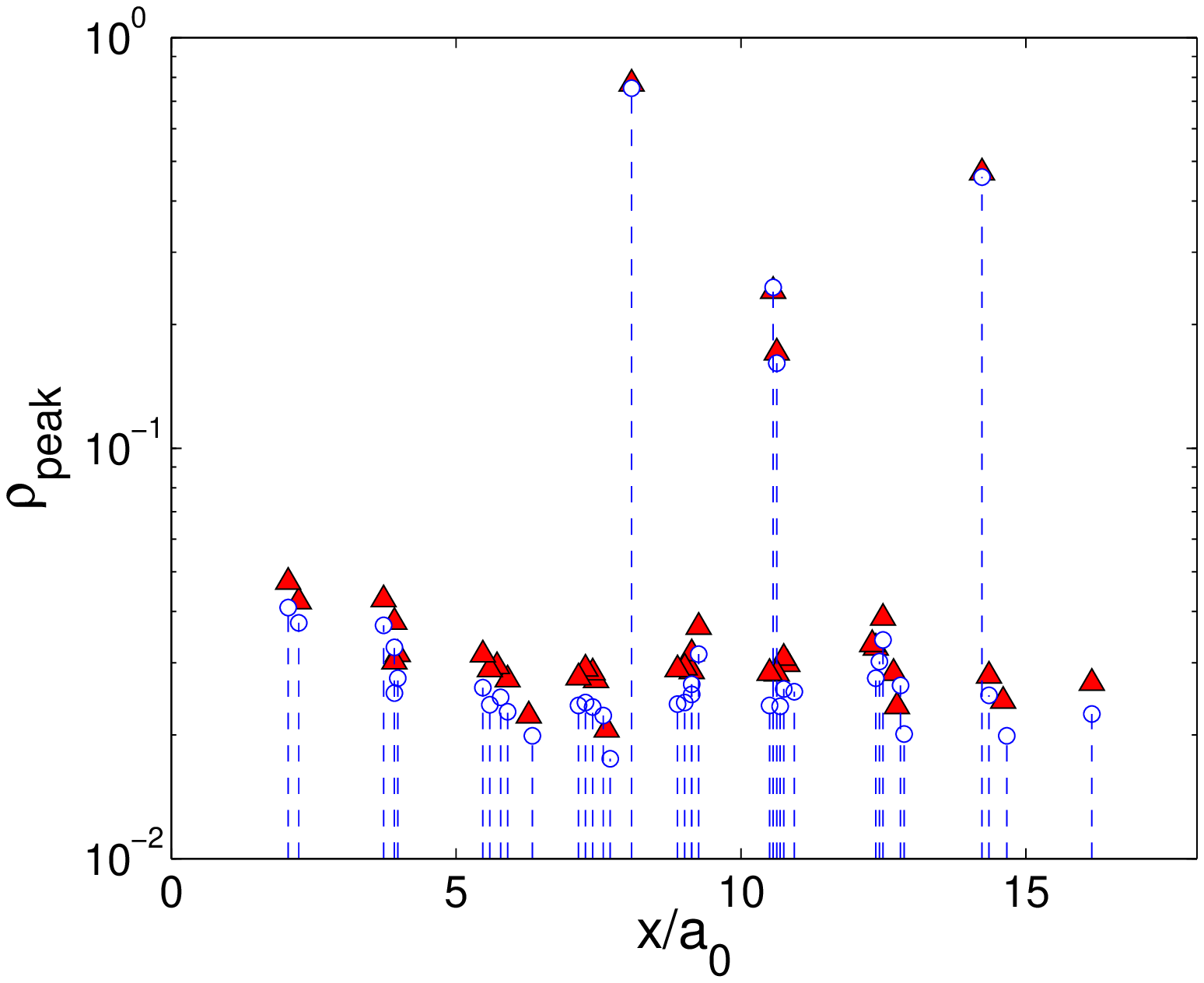}
\caption{ (Color online) Peak height vs position for $c=1/9$, comparing  vertical and tilted
($\tan \psi=0.437$) cases. Results at $\psi=0$ are shown as the (red) triangles and
those of the tilted case by the (blue) dots and impulses. The top panel is for $T=18.4 K$
and BoG states while the bottom one is at $T=17.8 K$ (BrG).}
\label{fig8}
\end{figure}

There are some additional 
noteworthy differences between the tilted pin results and the results
for vertical pins.
We have already seen that the
transition temperatures from BrG to BoG (Fig.~\ref{fig6})
and from Bog to IL (Fig.~\ref{fig7}) 
are  higher for vertical pins. 
In the Fig.~\ref{fig6} plots one can also observe
that the difference in the slopes at the crossing, which is a measure of the
latent heat per vortex, is smaller in the vertical pin
case, as compared to the tilted situation. 
An additional difference is plotted in
Fig.~\ref{fig8}. There we plot, in  a semilog scale, the local density peak height  
as a function of coordinate in the $x$ direction, for both the tilted case
(plotted with lines ending with dots) and the vertical one (triangles). This is done in one panel
at $T=18.4 K$ in the BoG phase and at $T=17.8 K$ (BrG) in the other panel. It is
striking that in both cases the peak heights for vertical pins are always
higher.

The free energy per vortex
is somewhat {\it lower} (at the same $T$) in the vertical case at lower values
of $T$ but the difference becomes negligible at sufficiently high temperatures
where the stable state is the BoG in both cases.  This occurs, we think,
for the following reason: in Fig.~\ref{fig8} and similar data, the integrated
vortex densities with values close to unity correspond to pin locations, indicating that the
pins are almost fully occupied by vortices. Therefore the portion of the free energy arising
from inter-plane electromagnetic interactions will tend to be higher in the tilted case.
However, Fig.~\ref{fig8} also shows that the smaller peak densities away
from the pinning columns are also higher for vertical pinning columns. This
means that the density distribution in the vertical case is more localized,
which is consistent with the higher transition temperature. At or above the
melting temperature of the pure vortex system without pins,
a more localized
density distribution will tend to have larger contributions to the free energy
arising from entropy and in-plane interactions. At temperatures  above $18.4 K$
the free energy gain arising from the lower localization in the tilted case
basically cancels the free energy cost from the inter-plane vortex interaction.

\subsection{Results at $c=1/8$}

\begin{figure}
\includegraphics[scale=0.5] {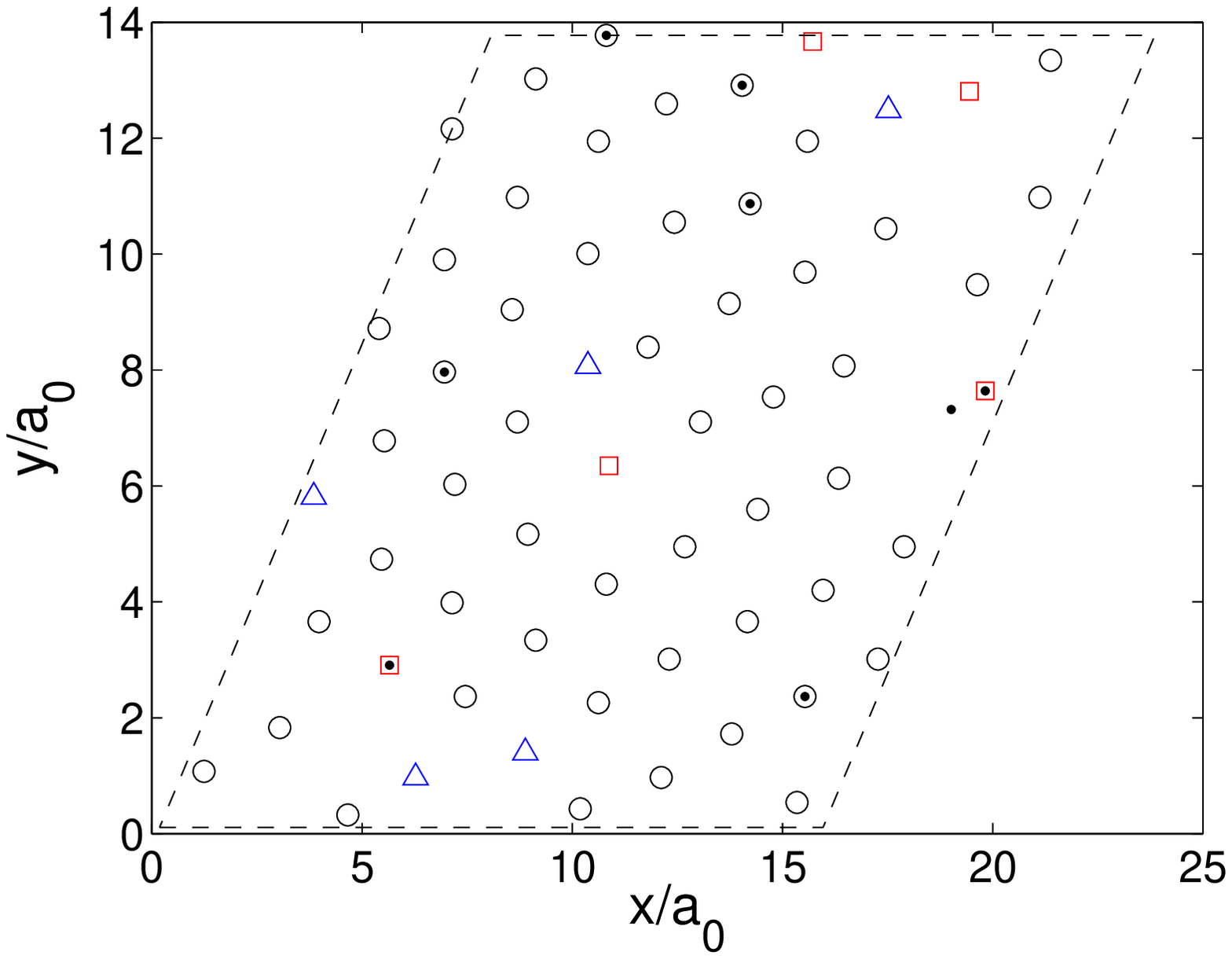}
\includegraphics[scale=0.5] {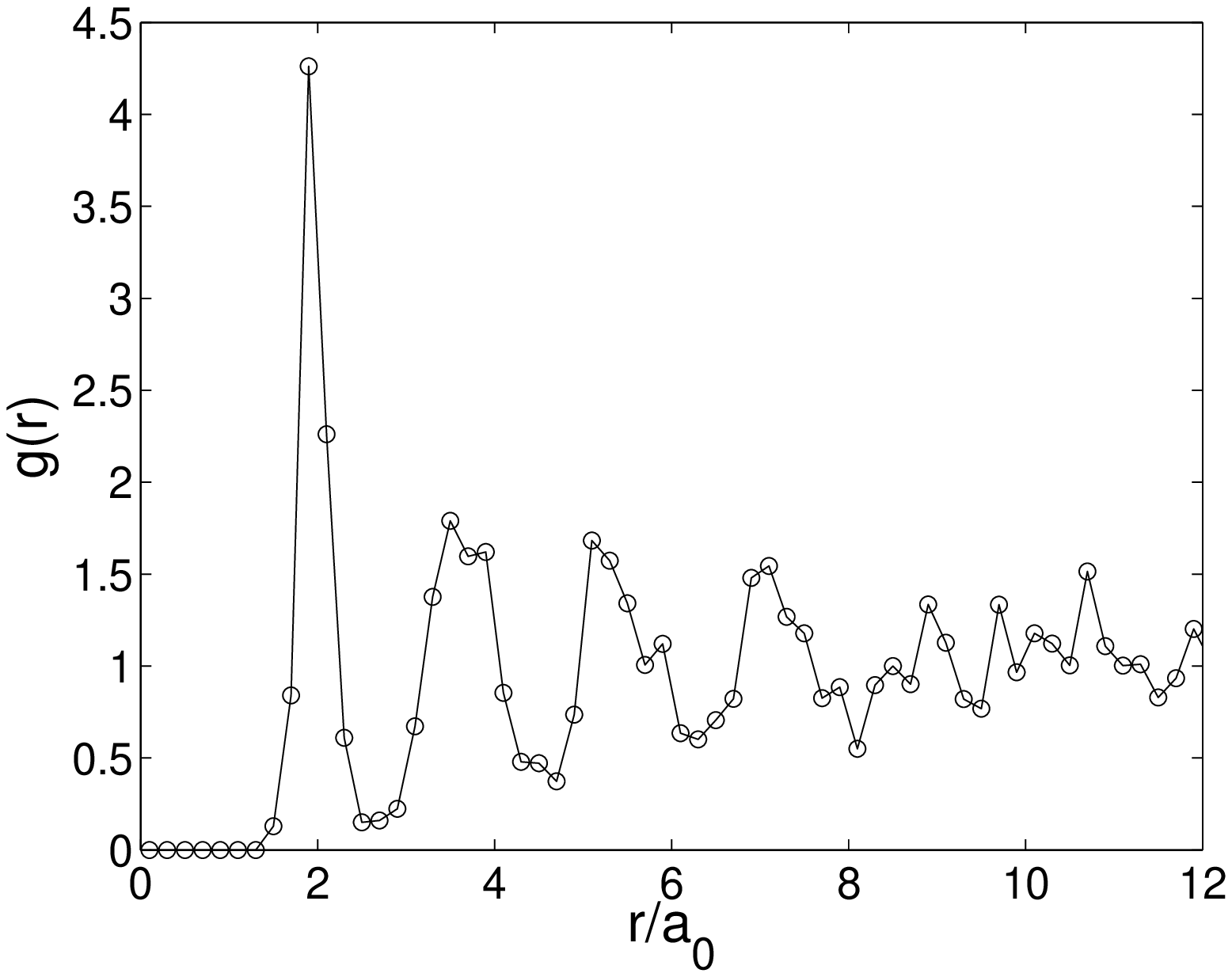}
\caption{Analysis of the structure of a BoG minimum for a tilted pin configuration at 
$N=128$, $c=1/8$, $T=16.8 K$. In the top panel we have shown the results of 
a Voronoi analysis. The symbols
mean the same as in Fig.~\ref{fig1}. In the bottom panel we have $g(r)$ plotted in
the same way as in Fig.~\ref{fig3}.}
\label{fig9}
\end{figure}

We have also studied a somewhat higher concentration, $c=1/8$ at $N=128$. This
corresponds to a somewhat larger tilt angle, $\tan \psi= 0.583$. Results for
the obtained equilibrium structure are given in Fig.~\ref{fig9}. The top panel 
of this  figure
shows the vortex lattice structure along with the results of a Voronoi analysis 
(completely analogous to Fig.~\ref{fig1}).
Despite the very low value of the temperature ($T=16.8 K$) we find that a good number
of defects remain and that the structure is the same as the BoG one in the top
panel of Fig.~\ref{fig1}. This is confirmed in the bottom panel of Fig.~\ref{fig9}
where we plot, at the 
same $T$, the correlation function $g(r)$ as in Fig.~\ref{fig3}. We see 
(compare with Fig.~\ref{fig3}) that
the correlation function structure is  of the BoG type. This remains the situation
down to the lowest temperatures reached ($T=15.2 K$). The free energy per vortex is not
too different from that in the $c=1/9$ case (Fig.~\ref{fig5}) but 
no instability to a more ordered state is found, down to 
the lowest $T$ attained. It is possible to obtain
BrG like structures
by quenching to low
temperatures with initial conditions corresponding to a crystal: 
we have done so by quenching to $T \le 16.0 K$  but the
resulting free energy values are considerably higher than those for the BoG at
the same $T$. Thus
in this case only the BoG
is found as an equilibrium state. 

We conclude that at these values of $c$ and $\psi$ no transition
to a BrG occurs except possibly at much lower temperatures. We have also studied the
same pin configuration, at this value of $N$, for vertical pins. We have again found no
BoG to  BrG transition upon cooling. We conclude then that the change in $c$, not the different
value of $\psi$, is responsible for the different behavior found in
the two cases studied here.
The high sensitivity of the possible BoG to BrG  transition to $c$  should not
come as a surprise. In previous work (see in particular Fig.~1 of
Ref.~\onlinecite{us3}) for vertical columns and a much larger
value of $N$, where because the problem is quasi two-dimensional
we were able to map the phase diagram in the $(T,c)$ plane
at constant field, we found that the line in
the $(T,c)$ plane separating the BrG from the BoG, while nearly vertical at 
small $c$, eventually curves sharply and then becomes nearly horizontal,
reflecting a very strong dependence of the transition temperature on $c$ and
leading in fact to the disappearance of this first order transition at somewhat
larger $c$.  This is quite consistent with what we find here.
There are however small quantitative differences 
with the results of Ref.~\onlinecite{us3}. Here,
we find that the BrG phase is still present at low temperatures for $c=1/9$, whereas 
Ref.~\onlinecite{us3} reported this phase  absent for $c > 1/32$. 
The transition and crossover 
temperatures found here are also slightly different from the values reported in our
earlier work. We believe that this
is due to the large difference between the sizes of the systems considered.
Since the system with vertical columnar pins is effectively two-dimensional, 
it was possible to study
much larger systems (with $N_v=4096$,  about 100 times larger than those
considered here) in our earlier studies. The smallness of the system size in
the present study makes the results quantitatively less reliable: this is clear from the observed
sample-to-sample variations of the transition and crossover temperatures.
Our earlier results  obtained for much larger samples, 
would be more reliable for vertical pins. 
The purpose of considering vertical pins in the present work was to make
a direct comparison with the behavior for tilted columnar pins without having to worry about
sample to sample variations or finite size effects.

\section{Summary and discussions} 
\label{summary}

Our detailed 
comparison of the results  for the thermodynamic behavior of the 
vortex system in the presence
of a dilute array of tilted columnar pinning centers 
reveals significant quantitative differences between this system
and a smilar system with  vertical pinning columns, normal to the layers, in the same
in-plane configuration.   
The thermodynamic behavior of the tilted pins system is 
however qualitatively similar to that
found in our earlier studies~\cite{usprl,us2,us3} of the vortex system with  
columnar pins perpendicular to the layers. In both cases, all the pins are occupied 
by vortices if the relative concentration $c$ of the pinning centers is small. In the
tilted case, we find that the interstitial vortices are well-aligned in the tilt direction.
If the relative pin concentration is low ($c=1/9$), the low-temperature phase exhibits the
characteristics of a Bragg glass. As the temperature is increased, this phase transforms, via
a first order transition, to a more disordered BoG phase which crosses over to
an interstitial liquid at a slightly higher temperature. For a higher pin concentration
($c=1/8$), the Bragg glass phase is absent and the system exhibits only the crossover 
from the low-temperature BoG to the interstitial liquid phase as the temperature
is increased. This is qualitatively similar to what occurs in the vertical pins case.

Quantitatively, 
the temperatures at which the
transition from the BrG phase to the BoG phase and the crossover from the BoG to the interstitial
liquid occur are found to be appreciably higher (by about one degree, or over 5\%) 
in the vertical pin case. The degree
of localization of the vortices in the low temperature, solid-like phases is also significantly
higher for vertical pins. We attribute these differences to the ``frustration'' in the tilted case, 
arising from a competition between the interlayer vortex interaction, which is minimized when the
pancake vortices on different layers are stacked in the vertical direction, and the pinning energy
which is minimized when the vortices are aligned in the direction of the tilt. This competition
also makes the free energy in the tilted case slightly higher than that for vertical pins at
low temperatures, as we have seen. 
These physical effects of tilting the columnar pins away from the layer
normal should be observable in experiments.


The differences we find between the results for vertical and tilted pins seem to contradict
some  experimental~\cite{zeldov07} studies
which concluded that the thermodynamic behavior of the vortex system is independent of the angle
between the magnetic field and the tilt direction if the areal densities of pancake vortices and 
pinning centers on each layer are held fixed. It is important to understand the reasons for this
apparent disagreement. In the experiment of
Ref.~\onlinecite{zeldov07}, the effects of changing the angle between the magnetic field and the 
direction of columnar pins were explored by changing the field direction for a sample with 
columnar pins tilted by
45$^\circ$ from the layer normal. This is not the same as the situation considered in our study.
In an isotropic superconductor, the individual directions of the field and the columnar pins are
not important: the behavior of the vortex system is determined by the angle between the two
directions. But for highly anisotropic layered materials such as high-$T_c$
superconductors, the directions of both the field and the columnar pins 
are important. The experiment of Ref.~\onlinecite{zeldov07} did not present any 
comparison between the results obtained for the two cases considered in our study:
one in which the pins are tilted away from the layer normal, and the other in which the pins
are perpendicular to the layers, but the areal densities of the pins and pancake vortices
on each layer are the same as those in the first case. Since the measurements for different 
orientations of the field were carried out for the same sample with the columnar pins tilted
away from the direction of the layer normal, the frustration effects mentioned above, 
arising from the competition between interlayer interactions and pinning, were present in all the
measurements. In contrast, these frustration effects are not present in one of the cases
(vertical pins) considered in our study. Thus there is no real contradiction.
In view of our results, an experiment that makes 
a comparison between the thermodynamic behavior in 
the two cases considered in our study would be very interesting.

It is more difficult to understand the reason for the difference between our results 
and those of Langevin simulations~\cite{zeldov07,gl07} performed on systems very similar to those
considered in our study. The simulations described in these papers were carried out
for both vertical and tilted columnar pins, keeping the areal densities of pinning centers and
pancake vortices fixed. Both electromagnetic and Josephson interactions between pancake vortices 
on different layers were included. Since both these interactions prefer
vortices on different layers to stack up in the direction of the layer normal, the frustration
arising from the competition between these interactions and the pinning potential for tilted
columnar pins is expected to be stronger in these simulations in comparison to that in our 
study which considers only the electromagnetic interaction. However, these simulations did not
find any significant difference between the results for vertical and tilted pins. This disagreement
with the results of our study may be a consequence of differences in 
system parameters. The values of $c$ used in the simulations ($c=0.35$ and $0.5$) are substantially higher
that those (1/9 and 1/8) considered here. A large concentration of pinning centers
has the effect of reducing the 
relative importance of the interlayer interactions by making the pinning energy the dominant
term in the total energy of the vortex system. In fact, it is argued in 
Refs.~\onlinecite{zeldov07,gl07} that the cost in Josephson and electromagnetic 
energies due to the tilting of the vortices is negligibly small compared to the gain in pinning 
energy for the parameters used in the simulations. If this is so, then it is not surprising that
the simulations did not find any difference between the thermodynamic behavior for tilted and
vertical pins. It is also possible that the  simulations are not 
sufficiently accurate to capture the 
fairly small differences between the results for the two cases
found in our study. The relatively small size of the simulated systems ($N_v=36$, 
and a number of layers $N_L=200$, 
which is substantially smaller than that considered in our study) implies that there 
would be large fluctuations in the quantities measured in the simulations. This would lead to
substantial uncertainties in the determination of transition temperatures -- it is well-known that
it is very difficult to determine transition temperatures accurately from simulations of
small systems. 
The authors mention in Ref~.\onlinecite{gl07} that 
their simulation is not accurate enough to determine transition temperatures with an accuracy of
$1 K$. Since the differences between the transition and crossover temperatures for tilted and
vertical pins found in our study are of the order of $1 K$, these differences would not be
detected in the simulation. 
Some of the detailed comparisons between the results for the 
two cases, shown in Fig.~8 of Ref.~\onlinecite{gl07}, are actually in agreement with our observations.
For example, it is shown in panel (c) of Fig.~8 of  Ref.~\onlinecite{gl07} that the mean-square displacement
of the vortices from their equilibrium positions is slightly higher in the tilted case. This is
very similar to the results shown in Fig.~\ref{fig8} above.
We expect that the other differences between the results for vertical
and tilted columnar pins found in our study will also be observed in simulations if the measurements
are done with sufficient accuracy at the same values of $c$ and other relevant parameters.

%


\begin{acknowledgments}
This work was supported  in part by NSF (OISE-0352598) and by
DST (India).
\end{acknowledgments}


\begin{thebibliography}{0}
\bibitem{old} Thus, we cannot compare with the different 
configurations from Ref.~\onlinecite{us3}, where 
a much larger $N$ was used. 
\end{thebibliography}

\end{document}